# Experimental study on hard X-rays emitted from metre-scale negative discharges in air

P O Kochkin[1], A P J van Deursen[1], U Ebert[1,2]
[1] Department of Electrical Engineering, Eindhoven University of Technology, POBox. 513, NL-5600 MB Eindhoven, The Netherlands
[2] Department of Applied Physics, Eindhoven University of Technology, and Centre for Mathematics and Computer Science (CWI), POBox 94079, NL-1090 GB Amsterdam, The Netherlands
E-mail: p.kochkin@tue.nl

**Abstract.**

We investigate the development of meter long negative discharges and focus on their X-ray emissions. We describe appearance, timing and spatial distribution of the X-rays. They appear in bursts of nanosecond duration mostly in the cathode area. The spectrum can be characterized by an exponential function with 200 keV characteristic photon energy. With nanosecond-fast photography we took detailed images of the pre-breakdown phenomena during the time when X-rays were registered. We found bipolar discharge structures, also called "pilot systems", in the vicinity of the cathode. As in our previous study of X-rays from positive discharges, we correlate the X-ray emission with encounters between positive and negative streamers. We suggest that a similar process is responsible for X-rays generated by lightning leaders.

1. Introduction

Thunderstorms are held responsible for Terrestrial Gamma-Ray Flashes (TGFs) – the most intensive pulses of electromagnetic radiation in the terrestrial atmosphere [1]. TGFs were first detected from space [2] and later at ground level [3,4]. The precise mechanism of their generation is still under discussion. The two most investigated theories are based upon relativistic feedback [5] with its continuation as "dark lightning" [6], and upon production of energetic electron at the tip of a lightning leader [7–10].
Moreover, X-ray bursts emitted by lightning leaders are an intriguing but as yet unsolved problem in lightning physics [11]. The X-rays bursts have been detected both from natural and from rocket-triggered lightning. In natural lightning X-rays were detected during the stepping process of lightning leaders [12], and they were later correlated with a single step [13]. A negative leader often steps by forming a space leader or space stem in front of it. The space leader is a bipolar structure that grows in both directions. The step occurs when the positive part attaches to the main leader. In triggered lightning X-rays originate from the tip of a "dart leader" [14] which also propagates in steps [15].

In the laboratory negative metre-long discharges can also grow through a space stem/leader formation [16–18]. And, as was first shown in [19] and later confirmed by several different high-voltage laboratories, long sparks also produce bursts of X-rays [20–23]. While Relativistic Runaway Electron Avalanches (RREAs) cannot be responsible for the X-ray emission in laboratory sparks [24], the thermal electron runaway mechanism provides a reasonable explanation for such emissions [7,9,25–29]. The thermal runaway electron mechanism relies on the assumption that some region with strong electric field is created by the discharge. The mechanism is briefly described in Section 3. In [7,9,27–29] it was shown that the tips of negative streamers can accelerate electrons into the run-away regime. Cooray et al. [30] suggested that the run-away effect might be enhanced between positive and negative streamer tips approaching each other. In our previous study of long positive laboratory discharges with nanosecond-fast photography we confirmed this suggestion [31].

In the present study we investigate source location, mechanism and characteristics of the X-ray bursts from *negative* discharges by measurements. In time-resolved photographs we show the space stem formation, the development into a pilot system and the attachment to negative leader/electrode.

## 2. Experimental setup

The setup is similar to that described in [31,32] and represented in Figure 1. Here we describe only the essential elements. A 2 MV Marx generator delivers a standardized lightning pulse with 1.2/50 μs rise/fall time when not loaded by the spark gap. The voltage is applied between conical electrodes, where the high-voltage electrode acts as cathode, and the grounded electrode as anode. We use three distances between the cone tips: 107, 145 and 175 cm. Only the first two lead to a full gap breakdown. The upper voltage limit is set to about 1 MV. Two Pearson 7427 current probes with 70 MHz bandwidth determine the currents through the high-voltage and the grounded electrodes. The probe for the high-voltage (HV) electrode has an optical transmission system to the oscilloscope. Appropriate attenuators and two antiparallel high-speed diodes protect the input of the transmitter. The diodes limit the linear response to 250 A. The grounded (GND) electrode current probe connected to the scope directly. Vaulted aluminum discs protect the probes against the spark current of 4 kA.

Two LaBr$_3$(Ce$^+$) scintillator detectors (D1 and D2) manufactured by Saint-Gobain are mounted in EMC-cabinets and record the X-rays. The quality of the EMC cabinets is such that the discharge formation does not interfere with the X-ray signals. The scintillator crystals are cylinders of 38 mm diameter and length, encased in 0.5 mm thick aluminum. The case is transparent for X-rays above 30 keV (attenuation 15% or less). The properties and performance have been discussed in [22,33]. The scintillators have a fast rise/decay time (11/16 ns) and a light yield of 63 photons/keV. The light is further amplified by a photomultiplier with special HV dynode dividers and additional capacitors between the upper dynodes to increase speed and to reduce saturation. The photomultiplier signal is measured with an oscilloscope, where we used the well protected high-impedance input, preceded by an external 50 Ohm cable termination. This termination limits the rise time to about 1 ns, better than needed for our experiment. Gamma spectrometers often apply wave shaping via differentiation and integration of the incoming signal pulse in combination with sample-and-hold units to allow a slow AD conversion. The amplitude accuracy is then high, at the cost of speed. In our experiment we valued time resolution over amplitude resolution, reason why we used a 10 GS/s 8 bit oscilloscope. The linearity of the detectors has been tested in [34] on $^{241}$Am, $^{137}$Cs, $^{60}$Co and remains better than 6% at least up to 2.5 MeV. At higher energies saturation of the photomultiplier causes a slight deviation from linearity. By averaging many signals of the 662 keV photopeak from $^{137}$Cs we obtained a noise free single photon response. The response as function of time allows distinguishing individual pulses even when pile-up occurs, as will be shown later in detail. The detectors are placed at different positions around the gap. To determine the origin of the X-ray signals in some experiments we restricted the detector field of view by 15 mm thick lead cylinders to a solid angle of 0.23 Sr, and we pointed them to different parts of the gap. Lead attenuators of different thickness helped us to determine the energy distribution.

The Picos4 Stanford Optics camera with an intensified CCD [35] is placed at 4 m distance from the spark gap and is directed perpendicular to the spark axis. The camera and all communication lines are properly shielded against electro-magnetic interference. The camera has a monochrome CCD; in contrast to [32] a single camera is employed. The lenses were either a Nikon 35 mm F2.8 fixed focus or a Sigma 70-300mm F/4-5.6 zoom.

The electrical signal acquisition system consists of two Lecroy oscilloscopes with 1 GHz bandwidth. The negative edge of the signal from the high-voltage divider triggers the oscilloscopes. One of the oscilloscopes transmits the trigger to the camera. The differences in the delays caused by the instruments and cables have been corrected to within ns accuracy.

## 3. Results

Anticipating on the results of Section 3, we first discuss the influence of the laboratory background on the x-ray measurements. We measured the number of detector pulses several times during minute long time periods, with the trigger level set at 30 keV and the detector placed in one of the positions near the spark gap without discharges. The average count rate was 50 per second. This includes the contributions from cosmic rays, the laboratory environment and the internal isotope decay. When the Marx generator fires the detector registers x-ray signals within a fixed time window of about 500 ns when the voltage is sufficiently large, as will be discussed in Section 3.2. The a priori probability to observe any background pulse in that window is equal to 500 ns / 20 ms = $2.5 \times 10^{-5}$. Moreover, we also observe discharges with two or three x-ray bursts. The probability that these are due to background is even smaller. However, the background might influence the attenuation measurements (Section 3.5) for attenuations of the order of a factor of 100. Several reports on the observation of x-rays from long laboratory sparks have been published [20–23]. Blank tests with a photomultiplier without scintillator [19] showed that the spark caused no interference on the signal. We have many spark measurements with only noise recorded on detectors with the scintillator present; these measurements might be considered as a more stringent test on the background.

We now briefly describe the process of electron run-away responsible for the X-ray production. Electron run-away was first described by Wilson [36] and later by Gurevich [25]. If free electrons are exposed to an electric field in ambient air, they will be accelerated in the field and lose their kinetic energy in inelastic collisions with air molecules, and in this manner they will approach some average drift velocity in the field. However, they can also get into the run-away regime, where they gain more energy in the field than they lose in collisions. For this to happen the electron need to reach energies above 100 eV; for this energy the momentum transfer collision frequency and hence the effective friction is maximal. The higher the electron energy above 100 eV, the lower the electric field required to maintain the runaway state, down to a minimum of 2 kV/cm for electron energies of ~1 MeV. It has been shown by simulations [7,28,29] that negative streamers can accelerate electrons into the run-away regime. These high energy electrons can generate X-rays by Bremsstrahlung when colliding with air molecules, according to a theory first described by Bethe and Heitler in 1934. These are the X-rays measured by our detectors.

*3.1. Discharge development*

The development of the negative discharges has been characterized in detail through ns-fast photography in [32]. For the sake of completeness we briefly recall the processes here. The pulses of the high-voltage current shown in Figure 2 indicate that the discharge development in a 107 cm gap can be divided into seven stages. All stages are very reproducible for different discharges with the same electrode configuration, and the curves in Figure 2 are actually averages over 302 discharge pulses. The first four stages coincide with four bursts of negative streamers. When we apply the high voltage, a negative streamer corona appears near the HV electrode and develops downwards and horizontally. It extends until the ratio of instantaneous potential over length is about $E_{min}$=12 kV/cm, which is the so-called stability field for negative streamer propagation; a new critical discussion and test of the stability field concept can be found in [37]. As a negative streamer cannot propagate slower than the electron drift velocity at the enhanced electric field at the streamer tip [38], it grows with a velocity of at least $5 \cdot 10^5$ m/s to the length determined by the instantaneous voltage. Meanwhile the voltage continues to rise, and eventually a second burst of negative streamers is emitted from the HV electrode and propagates further into the gap. The four stages of development correspond to four corona and streamer burst, each one propagating further into the gap, while the voltage rises. In addition, the inductive impedance caused by the long wires between HV divider and gap also contribute to limiting the current rise.

Stage 1 corresponds to the formation of a negative inception cloud around tip and protection disk of the high-voltage electrode. The inception cloud destabilizes and ejects negative streamers [39]. Stages 2, 3 and 4 correspond to the second, third and fourth streamer burst, respectively. They appear for all gap lengths larger than 1 m. When the outermost negative streamers approach the grounded electrode they bring part of the high voltage downward and enhance the local electric field there which, in turn, leads to the formation of a positive inception cloud on the tip and on the sharp edges of the grounded electrode. The positive counter-streamers emerge from the positive inception cloud and move upwards, where they merge with the negative streamers. For the 107 cm gap the outermost streamers cross the gap at the fourth burst and then create a spark. In stage 5 the positive streamers from the grounded electrode reach to the high-voltage electrode, possibly along the traces of previous negative streamers. Then high amplitude HF oscillations in the cathode current may occur. A conductive channel between the electrodes is established and the currents on both electrodes increase quickly; this is the beginning of the leader phase (stage 6) and of complete breakdown (stage 7). It might be useful to recall at this point, that streamers emit light only in their growth region where additional ionization is created, and not in their conducting, current carrying parts. Only the high current in a leader or spark can create an optical signal; it can be distinguished spectroscopically from streamer heads.

*3.2. Influence of the gap length on the electrical characteristics and on the X-ray time*

In Figure 3 (a) to (c) we compare the electrical characteristics and X-ray emissions from gaps of 107 and 147 cm length, both at a maximum voltage of 1.1. MV. All curves are averaged over 302 or 72 discharges, respectively. The electrical characteristics of the individual discharges at the same gap length are so similar, that the averages show essentially the same features as the single measurements. The voltage rise time is determined by the Marx generator and the HV divider circuit. Because the discharge current is strongly determined by the negative corona development and by the high inductive impedance of the wire between the top of the HV divider and the high-voltage electrode, the cathode current up to 1 μs is remarkably independent of the gap length. Most X-rays appear between 0.65 and 0.9 μs, for both gap lengths (Figure 3 (d)); this time interval largely coincides with the fourth streamer burst. The voltage is then over 500 kV which apparently suffices to accelerate electrons sufficiently that their bremsstrahlung photons are within the energy range of our detectors. That the X-rays are detected within the same time span for both gap lengths implies that the anode region cannot contribute much to the X-ray generation. It takes 2 μs longer to break the 147 cm gap down than the 107 cm gap. As a result, the anode current also rises much later in the 147 cm gap, and about 1.5 μs after the X-ray detection. At breakdown the voltage fall time is determined by the resonance frequency of the capacitive HV divider and the inductive gap circuit.

In the 147 cm gap, the streamers are not able to cross the gap at the fourth burst and a dark period without anode current occurs between 1.1 to 1.6 μs in Figure 3 (c). A fifth streamer burst does not occur anymore because the voltage does not rise anymore. There is no optical activity in the gap (the dark period) until positive streamers develop near the grounded anode. At this moment we see the current through the anode rising rapidly, as shown in Figure 3 (b) at about 2 μs. Although there is hardly any light during the dark period, this does not indicate that there is no current flowing, but only that hardly any ionization reactions occur.

The high-frequency oscillations of the cathode current in Figure 3(c) at t = 1 to 1.1 μs correspond to the moment when positive streamers from the grounded electrode collide with the high-voltage electrode in the 107 cm gap. The same oscillations are visible at t = 2 to 2.1 μs for the 147 cm gap. Sometimes they are also accompanied by an X-ray signal.

*3.3. X-ray measurements*

A typical oscillogram with X-ray detection is shown in Figure 4. The gap distance between the electrodes is 107 cm. The voltage rises from 10 to 90% of its maximum value of 1.1MV within 0.7 μs, and breakdown occurs 1.6 μs after the start of the voltage pulse. In this measurement both X-ray detectors were placed next to each other at position H (Figure 1) with a centre to centre distance of only 6 cm as shown in Figure 5. When the X-rays are detected, in 82% of the cases the signal appears as a single pulse on one or both detectors simultaneously. In 17.5% of the cases we detect two X-ray pulses, well resolved in time during one discharge. And in the remaining 0.5% we detect three X-ray pulses. As shown in Figure 4, the X-ray signals appear simultaneously on both detectors. We conclude first that the X-rays are generated within nanosecond bursts, and secondly, taking into account that the scintillators have a diameter of 38 mm, that photon pile-up may occur in each detector. Still, all measured X-ray signals up to 0.5 MeV can be fitted with a single photon response. With a slight deviation from linearity, this is also possible for 2 MeV energy deposited in the detector; see for example Figure 6(a). This 2 MeV signal can only be explained by pile-up since the maximum of the applied voltage is 1.1 MV, and since ionization with two elementary charges (2e) is negligible. The rising slope of the signal indicates that all X-ray photons arrived within 6 ns. Even much larger deposited energies occurred, as is shown in Figure 6(b), where the oscilloscope channel clipped at its maximum of 5.5 MeV. The recorded detector signal can be fitted by two single photon responses with a delay of 40 ns, scaled to 10.3 and 7.5 MeV, respectively. However, as discussed in [34] deviations from linear response due to the saturation of the photomultiplier set in at 2.5 MeV. So the large signal may be additionally broadened due to different arrival times of the X-rays or by saturation. Both effects are difficult to distinguish, even in a non-clipped registration of the wave shape.

*3.3.1. Correlation between X-ray bursts and high-frequency oscillations of the current*

As in positive discharges (see Figure 4 in [31]), the X-ray signals are accompanied by high frequency oscillations of the cathode current. In Figure 7 such oscillations are marked by arrows. They are also visible in Figure 4 as sharp spikes on the HV-current curve at the moment of X-ray detection. The more pronounced the oscillations are, the more likely they are accompanied by X-rays and the higher the amplitude of the X-ray signal is. As mentioned above, we do not detect X-rays in 100% of the discharges, but the oscillations are clearly visibly in every discharge. Since the X-rays come in short bursts and are associated with high frequency oscillations, we assume that a ns-fast process is responsible for their generation, for instance the encounter of positive and negative streamers. This can happen at least three times (three X-ray bursts) during one discharge. As the X-rays appear in bursts, it is unlikely that the continuous propagation of streamers or leaders drives the process. However, any sudden process like stepping or collision is a candidate.

*3.4. X-ray registration rate*

By comparing the X-ray detections for different detector positions, we may derive where the X-rays are generated. Table 1 shows for each detector position the ratio of the number of discharges with X-ray detection over the total number of discharges. The gap between the electrodes is fixed at a length of 107 cm. The data for positions A, B, C and D were obtained in one series of measurements. Those for E, F and G were obtained in another series two months later. Those at position H were measured even later. For the detector at positions A and B we placed a small EMC cabinet under the grounded electrode (as shown in Figure 1); this cabinet remained there during all measurements in series I and II. In contrast to our previous study of positive discharges [31], the cabinet did not influence the X-ray registration rate. This agrees with the observation that the grounded electrode essentially does not contribute to the X-ray generation in the present experiments. When two detectors were placed in different positions during one discharge (for example detector D1 at position A and detector D2 at position D) they often show x-ray signal simultaneously.

The measurements with the collimated detectors "F' up" and "F' down" indicate that 2/3 of the X-rays come from the upper half of the gap. Besides that, the registration rates at positions A and B are similar or lower than those at positions D, E, F and G. Since they are all located at approximately the same distance from the cathode, the X-ray emission is on average isotropic in these experiments. To substantiate this further, we now assume that the X-ray source is point like and has a constant luminosity. With only geometrical decay, the registration rate should follow the inverse square law. For such a source, we can find its location based upon data of Table 1 by fitting the observed occurrences at detector position $\vec{r}$ to the function $P(\vec{r})$:

$$P(\vec{r}) = \frac{P_0}{|\vec{r} - \vec{r}_0|^2} \qquad (0.1)$$

where $P_0$ is the source amplitude or initial occurrence, and $\vec{r}_0$ its position vector. All detector positions in Table 1 were used; those for position F without collimator. The best fit $\vec{r}_0$ is indicated in Figure 8 by a star. The ellipse around the star represents the 95% confidence bound. Although R-square goodness of fit is low (60%), the location of the source is in accordance with the measurements with collimated detector at position F. It is remarkable that the source is off axis. This can have several reasons. First, the X-rays do not come from one fixed point $\vec{r}_0$ in space. Second, and more importantly, the X-ray bursts from each electron acceleration event are not distributed isotropically, but are beamed in the direction of the main electron acceleration. A further investigation of the opening angle of such X-ray bursts and a reevaluation of the data is currently under way.

*3.5. Nanosecond-fast photography of the cathode region during X-ray emission*

The previous section demonstrated that the X-rays appear only during pre-breakdown. The majority came from the cathode region. In addition, the X-rays are correlated with high frequency oscillations of the cathode current, and their source location is near the cathode. For these reasons we pointed a nanosecond-fast ICCD camera to the vicinity of the cathode.

*3.5.1. X-rays without final breakdown of the gap*

When we increase the gap length to 1.75 m, no spark develops within the 50 μs of high voltage delivered by the Marx generator in the absence of electric breakdown. The electrical characteristics together with an image with 100 ns exposure time are shown in Figure 9; clearly the anode current – if there is any – remains hidden in the noise in the measurement. However, both scintillation detectors simultaneously register an X-ray signal at the same stage and time as in the smaller gaps. The camera shutter was opened just after the X-rays had been detected. As can be seen from image (b), there is streamer/leader activity around the high-voltage electrode. We conclude again that the grounded electrode is not directly involved in X-ray production.

*3.5.2. The vicinity of the cathode during X-ray registration*

Figure 10 shows the images for six different discharges, zooming in into the region just below the high-voltage electrode, at the moment when most X-rays are detected. All images have an exposure time of 50 ns, and the shutter in all images of Fig. 10 and in Fig. 9 (b) has been opened with the same delay after the beginning of the voltage rise (time zero in all plots). The images (a) to (f) have intentionally been ordered such as to illustrate the discharge development. Negative streamers (ns) leave isolated dots or beads behind during the propagation (image (a)). Later, the beads act as starting points for positive streamers (ps) (image (b)). We call these features positive streamers because they look like streamers, their velocity coincides with the velocity of positive streamers in our setup ($2·10^6$ m/s, see details in [31,32]), they move towards the negative high-voltage and they branch in this direction. Remarkably, the upward moving positive streamers co-exist with negative streamers that move downwards. Later positive and negative streamers collide (images (c) and (d)). X-rays are detected in discharge (a) 50 ns after the image; discharge (c) 40 and 110 ns after the image; and (d) 300 ns after the shutter was closed. Even when the streamer encounter is clearly visible on the image, this does not guarantee X-ray detection. And vice versa – when we detect an X-ray signal, the streamer encounter, which is possibly responsible for it, might not be located in the camera field of view. Apart from that, it is a matter of luck to take a snapshot – point the camera to the right place, open its shutter at the right moment and keep it open as short as possible. Even though we cannot link a single encounter with a single X-ray burst and prove their correlation, similar collisions between positive and negative streamers have been observed in positive discharges, also simultaneously with X-ray registration.

The entire structure that eventually develops out of the negative streamers, beads and positive streamers is a pilot system in the nomenclature of [18,40]; a schematic representation of the pilot system is shown in Figure 11. Images (e) and (f) in Figure 10 show that such pilot structures are common features in negative discharges in the laboratory. The structures were also observed in other experiments [16–18]. In larger gaps of a few meters, the pilot system can even develop into a space leader. In the center of images (d) and (e) two of these structures are clearly visible, but more than ten were counted in image (f).

*3.5. Energy spectra and attenuation curves*

In order to get a statistically meaningful X-ray spectrum, we analyzed the amplitudes of 636 X-ray signals collected with a single detector at position H in Figure 1. We found that the X-ray energy depends neither on the instantaneous voltage nor on the current. Thus, they are apparently generated by independent events. So, when two or three X-ray bursts were detected during a discharge we count them separately. As we have previously shown, even a small deposited energy could be the sum of several photons. Because of multiphoton registration and overlapping X-ray bursts, we can only get a pseudo-spectrum with our detector. Such a spectrum is shown in Figure 12, where we divided the energy scale in bins of 55 keV. As mentioned above, detected energies of up to 0.5 MeV can be fitted by a single photon response. Energies larger than this value are *more likely* multiphoton registration and/or burst overlap. Up to about 0.5 MeV the pseudo-spectrum can be fitted by an exponential function dn/dε ~ exp(-ε/$ε_c$) with a characteristic energy $ε_c$ = 0.2 MeV. This $ε_c$ agrees well with the energies reported in [21,31]. The average deposited energy over the entire spectrum is 0.55 MeV. So, on average we detect 2 – 3 X-ray photons by our detector per burst. If we assume that the X-rays within one burst are distributed isoptropically, we get approximately $10^5$ photons per burst over the complete solid angle of 4π.

In order to get more information on the distribution of the single photon energies we performed a series of measurements with lead attenuators in front of the detector. One detector was mounted in the small EMC cabined located below the grounded electrode at position A, the other in the large cabined at position D (see Figure 1). The detectors were wrapped in a 15 mm thick lead cylinder and the scintillator crystals were covered by lead caps with varying thicknesses of 1.5, 3, 4.5, 6 and 7.5 mm. Each cap was placed right in front of the scintillator, touching it. For each cap thickness we determined the X-ray detection in 50 discharges. The data are shown in Table 2. The detections without attenuator indicate the initial intensity of the source at the specific location. As expected, the amount of detected X-rays decreases with increasing cap thickness.

If for the moment we neglect multiphoton registration and burst overlap, the attenuation (removal) of photons from the initial burst as they pass through the attenuator would follow the equation:

$$\frac{I}{I_0} = e^{-\mu(\varepsilon) \cdot x}$$

where $I_0$ is the initial source intensity at energy ε, *I* the intensity after the lead attenuator, $\mu(\varepsilon)$ the linear attenuation coefficient at energy *ε*, and *x* the lead thickness. The attenuation coefficient $\mu$ is the sum of individual attenuation coefficients for photoelectric absorption and Compton scattering:

$$\mu = \mu_{ph} + \mu_{comp}$$

We can neglect Rayleigh scattering and pair production in our energy range. Moreover, since we put the lead caps right in front of our detector, and since the attenuator thickness is significantly thinner than the detector size, many Compton scattered photons will penetrate into the scintillator and interact with it. We also proved experimentally in [31] that we can neglect Compton scattering in our setup and only consider the photoelectric absorption inside the lead attenuator.

Taking the detector's quantum efficiency at energy ε into account, we have the following general relationship:

$$I = \int_{\varepsilon_{min}}^{\varepsilon_{max}} \eta(\varepsilon) \cdot I_0(\varepsilon) \cdot e^{-\mu_{ph}(\varepsilon) \cdot x} d\varepsilon$$

where $I_0(\varepsilon)$ is the initial source intensity at energy ε, $\eta(\varepsilon)$ is the quantum efficiency of the detector (about 100% at our energy range), and the absorption function $\mu_{ph}$ is taken from NIST [41]. Now we can calculate the attenuation curves for different detector positions. In Figure 13 we compare the measured attenuation curves (dashed lines) with those calculated under the assumption that single photons are registered (solid lines). The measured attenuation curves are below the ones calculated, which confirms that the real X-ray spectrum is softer than that indicated by the detectors. This means that in each burst we detect several lesser energetic photons simultaneously rather than one single hard X-ray photon. A monoenergetic X-ray beam of 0.2 MeV photons would undergo attenuation similar to the one measured at point A.

In this analysis we neglected the attenuation by 2 m of air since the additional total attenuation is only 3% at 200 keV and 8% for 30 keV which is the lower limit of the detectors. But a more thorough investigation is needed. In a Monte Carlo approach we start with electrons at energies between 0.1 and 1 MeV, calculate the x-ray production, and include the transfer of the electrons and x-rays through the air and the processes in the scintillator. The results will be presented in a future paper.

## 4. Discussion

Although negative and positive laboratory discharges possess similar features – streamers, leaders, counter-streamers and counter-leaders – their development is quite different. The photography with nanosecond fast cameras shows that laboratory discharges with positive polarity grow in a more continuous way [31]. Negative discharges have a more complex structure and development mechanism [32], in particular, they do not propagate continuously in the present set-up, but in four streamer bursts, and they form space-stems ahead of the negative streamers/leaders – at least in the fourth burst near the cathode. The X-rays from both discharges appear in short bursts. The measurements with the LaBr$_3$ detectors fix the upper limit of the burst duration at 6 ns for signals of up to 2 MeV. Other measurements with the faster BaF$_2$ [34] and with plastic detectors show that the bursts likely last as short as 1 ns. This made us look whether the images contain indication of such fast processes, which can be held responsible for the X-rays. The best candidate is the encounter of streamers. Streamer heads of both polarities are observed simultaneously near the cathode. The streamers move with approximately 2 mm per ns in these images; the measured diameter is 2 to 4 mm.

Models suggest that the electric field in front of a streamer is about 100 – 160 kV/cm [42]. When two long streamers of opposite polarity approach each other, the electric field between their tips dramatically enhances. This mechanism is suggested by Cooray et al in [30] and in general is confirmed here and in [31] by time resolved photography with simultaneous X-ray measurements. Note, however, that in positive discharges, the encountering streamers are primary streamers propagating through non-preionized air near the grounded electrode, while in the negative discharges they are near the HV electrode pre-treated by several earlier streamer bursts.

An electric field enhancement can also occur when streamers approach an electrode, in the so-called proximity effect. In this intense field the electrons from the negative streamer or electrode can overcome the friction barrier at ~100 eV kinetic energy over a fraction of a millimeter and run away [43]. The photographs in Figures 9 and 10 show that such encounters occur near the cathode, so there is ample voltage difference left with respect to the environment for the electrons to attain large energy, if the electrodes are not screened by plasma. In our setup with voltages over 500 kV, electrons can accelerate into the relativistic regime, and lose their energy rapidly through X-ray bremsstrahlung. It should be noted that much of our understanding of streamers relates to the first streamers propagating through virgin air, while in the present experiments, the fourth streamer burst produces most X-rays. Inhomogeneous background ionization created by the previous streamer bursts can create the bead structure near the cathode [44] that can be seen in the images.

X-ray bursts in laboratory discharges of both polarities are accompanied by high-frequency oscillations; see [31] and this work. The frequency is far above the 70 MHz working range of the current probes. We attribute these electrical signatures to the electrodes acting as oscillating antenna excited by the sudden current changes caused by the streamer encounters. Perhaps even the conductive streamers may act as such an antenna. Although we did not register X-rays from every discharge, we always observed the oscillations. Moreover, the oscillations coincide with X-rays if detected. In addition, the amplitudes of the oscillations are positively correlated with the probability of X-ray detection. We therefore presume that X-rays occur in nearly all measurements and that the detection is a matter of probability given by the limited number of photons and energy limit of our detectors 30 keV. Another possible explanation of the lower X-ray detection rate is that the X-ray bursts have a finite opening cone, and that X-rays are detected only if the detector is in the cone. Investigations of this question are now under way.

In short gaps of 1 meter or less, the X-ray bursts coincide with the onset of current at the grounded electrode, as is for example shown in Figure 2. High-frequency oscillations are then also visible in the anode current. Table 1 for the position "F' down" shows that about one third of the X-rays occur in the anode region. We again attribute this to the encounter of negative streamers with positive streamers, but now near the anode.

The fact that the final breakdown is not necessary for the X-ray production allows us to compare the negative laboratory discharge with the X-rays produced by a negative stepped lightning leader. It has been shown that natural negative lightning generates X-rays during the stepping of the leader [12,13]. The stepped leader propagates by creating space leaders. The space leader is a bipolar structure that develops in both directions in front of the lightning leader. When the positive part of it connects to the negative leader the step occurs. Recent high-speed video observation of the stepping process [15] allows us to suppose that X-rays from negative stepped leaders can be generated in the same way as described in this manuscript. The pilot systems observed in this work develop into space leaders in longer gaps and in natural lightning.

The attenuation curve shows that most large energy signals are due to pile-up. However, the fact that we still detected an X-ray signal behind 7.5 mm of lead indicates that a high energy tail exists in the electron and X-ray distribution. Photons of 200 keV have a chance of less than 1% to pass the attenuator; at 500 keV it is more than 10%.

It will be an interesting experimental task to create a single streamer encounter under controllable conditions and with sufficient energy pumped in.

## 5. Conclusions

Based upon observations of more than three thousand long negative laboratory discharges we arrive at the following conclusions:
- Nanosecond-fast X-ray bursts happen during the pre-breakdown process; the final breakdown of the gap is not necessary.
- It is most likely that streamer encounters are responsible for the X-rays, because the field enhancement between streamers tips makes it easier for the electrons to run away (thermal run-away mechanism [25]).
- Since lightning leaders propagate in a similar stepped manner, we propose that streamer encounters are responsible also for the X-rays from the leader.
- The X-ray spectrum in our measurements can be approximated by an exponential distribution function with a characteristic energy of about 200 keV. In order to calculate the precise spectrum and the initial number of high-energy electrons, simulations would be required. These simulations should include the relevant processes and many details of the setup.
- The X-ray bursts seem to be correlated with current oscillations, but current oscillations occur also without X-ray detections. A possible explanation is that the X-ray bursts have a finite opening cone and that the detector is not always inside the cone.

## 6. Acknowledgments

PK acknowledges financial support by STW-project 10757, where Stichting Technische Wetenschappen (STW) is part of the Netherlands' organization for Scientific Research NWO.

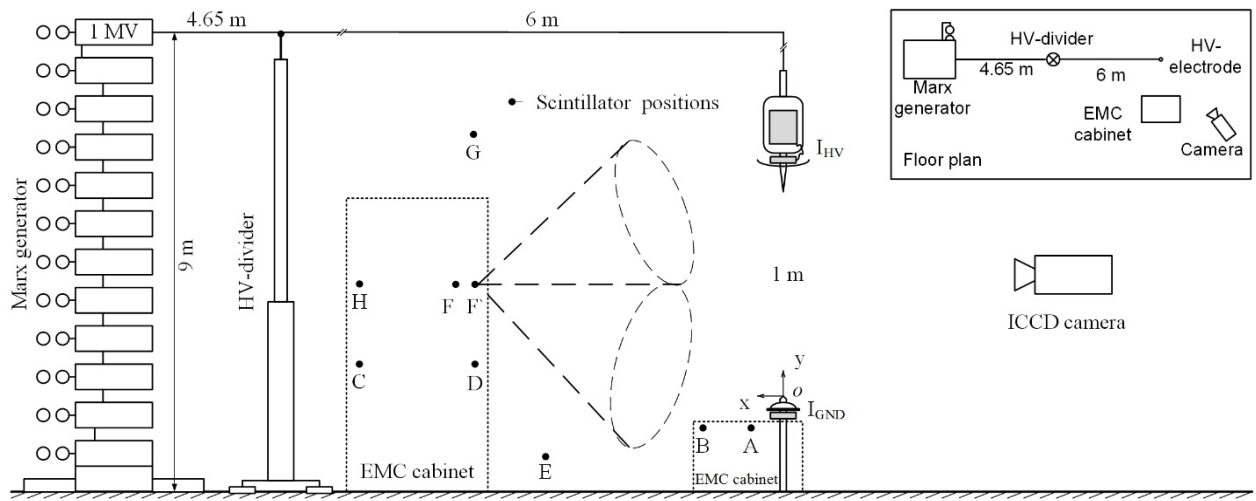

**Figure 1**. Schematic of the spark gap geometry. The positions of the X-ray detectors are labeled from A to H. They are all in the same vertical plane. The dashed cones indicate the field of view of the detector when it is placed inside a lead cylindrical collimator (see Section 3.4). The ICCD camera is located at a distance of 3.5-4.5 m from the gap. The distance between the Marx generator and the spark gap is approximately 10.65 m. The upper right inset shows the correctly scaled floor plan.

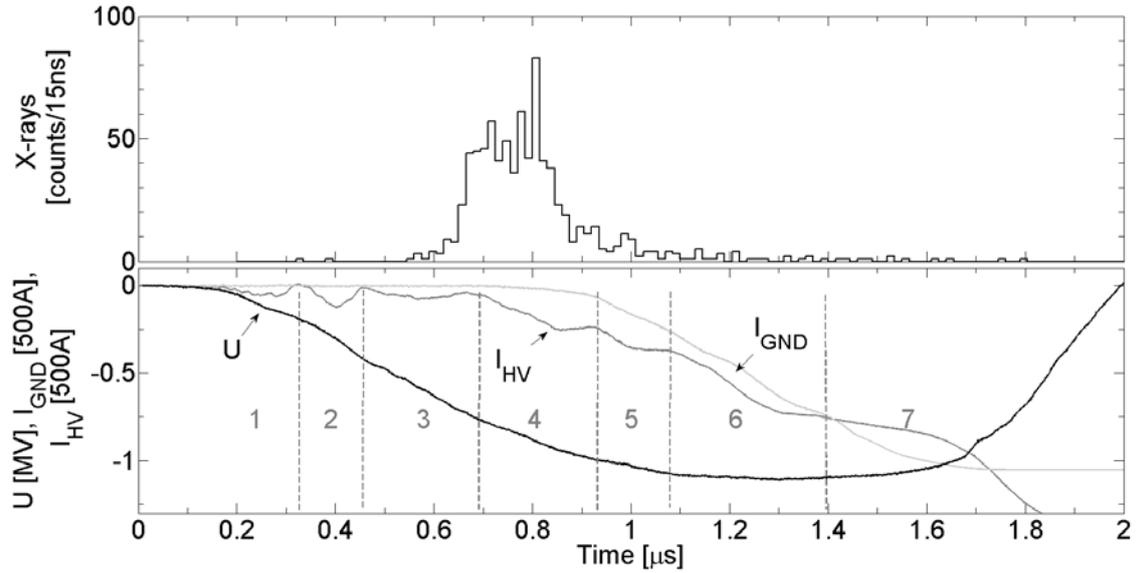

**Figure 2**. The development of the discharge in a 107 cm gap can be divided into 7 stages. Each stage begins with a rise of the current at the high-voltage electrode and ends with its drop. Voltage and current in the plot represent an average over 302 discharges. The X-ray counts per 15 ns represent data of 815 X-ray bursts detected during these 302 discharges. The maximum of the X-ray counts occurs at the beginning of the fourth stage.

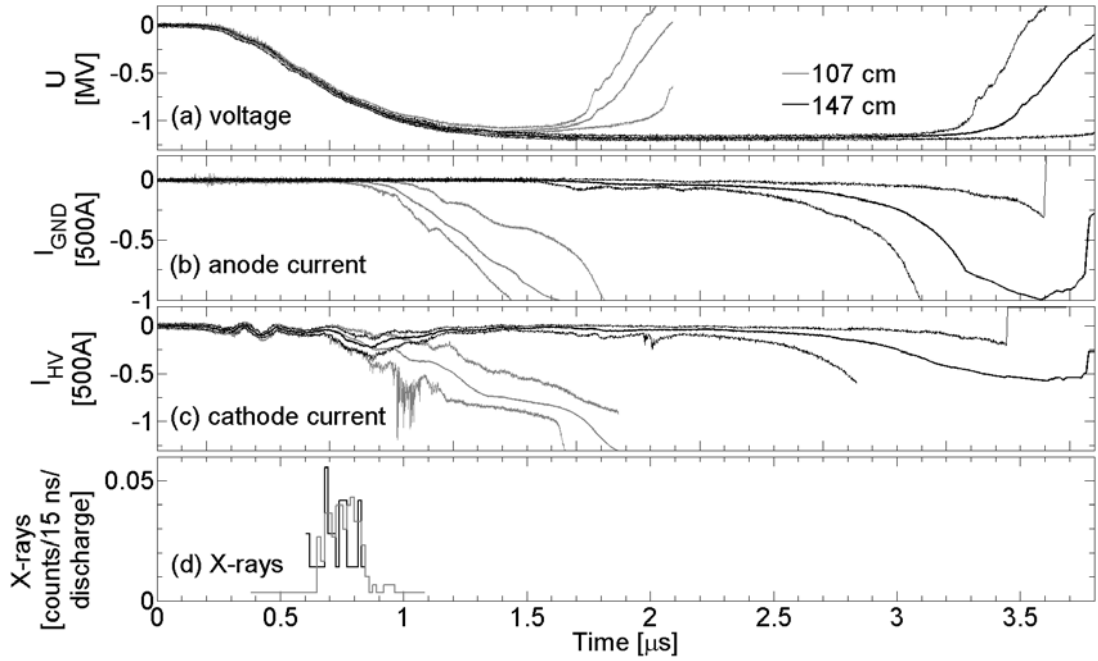

**Figure 3**. Electrical characteristics and X-ray registration time for discharge gaps of 107 cm (gray) and 147 cm (black). The measurements are averaged over 302 or 72 discharges, respectively. The electric breakdown in the 147 cm gap takes 2 µs longer than in the 107 cm gap. The cathode current curves (c) are remarkably similar up to 1 µs, and also the temporal distribution of detected X-rays is similar for both gaps where most emissions also occur during the initial 1 µs.

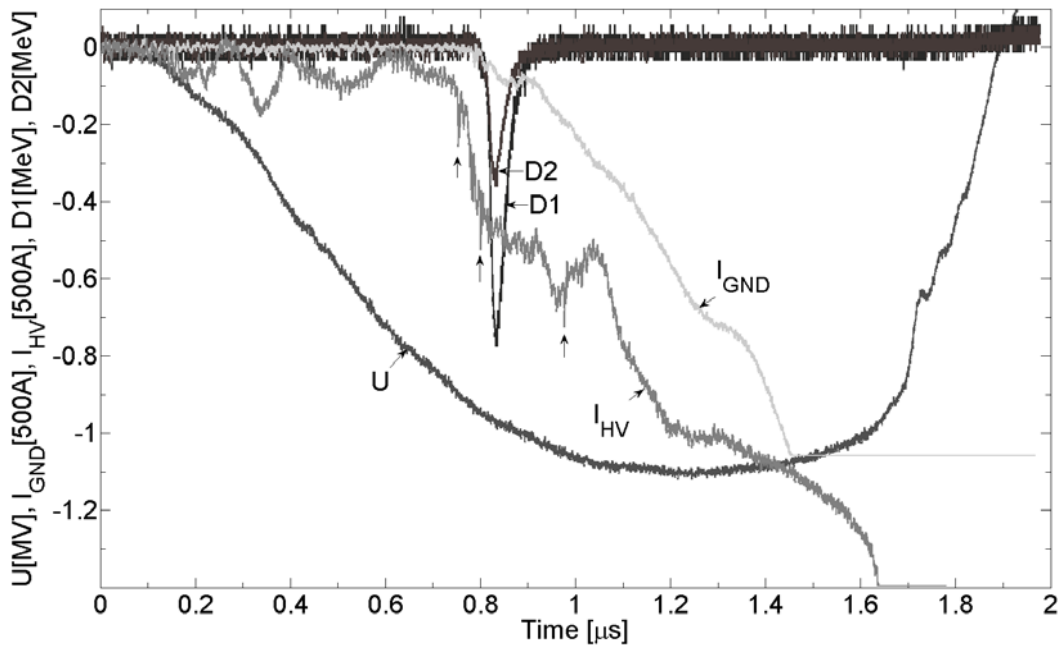

**Figure 4.** A typical recording of a single negative discharge. The voltage rises to 1.1 MV. The gap distance is 107 cm. Two LaBr$_3$ scintillation detectors D1 and D2 are placed next to each other at position H at 2 m distance from the spark gap. The X-ray detection coincides with the rise of the current I$_{GND}$ on the grounded electrode. HF oscillations are indicated by vertical arrows.

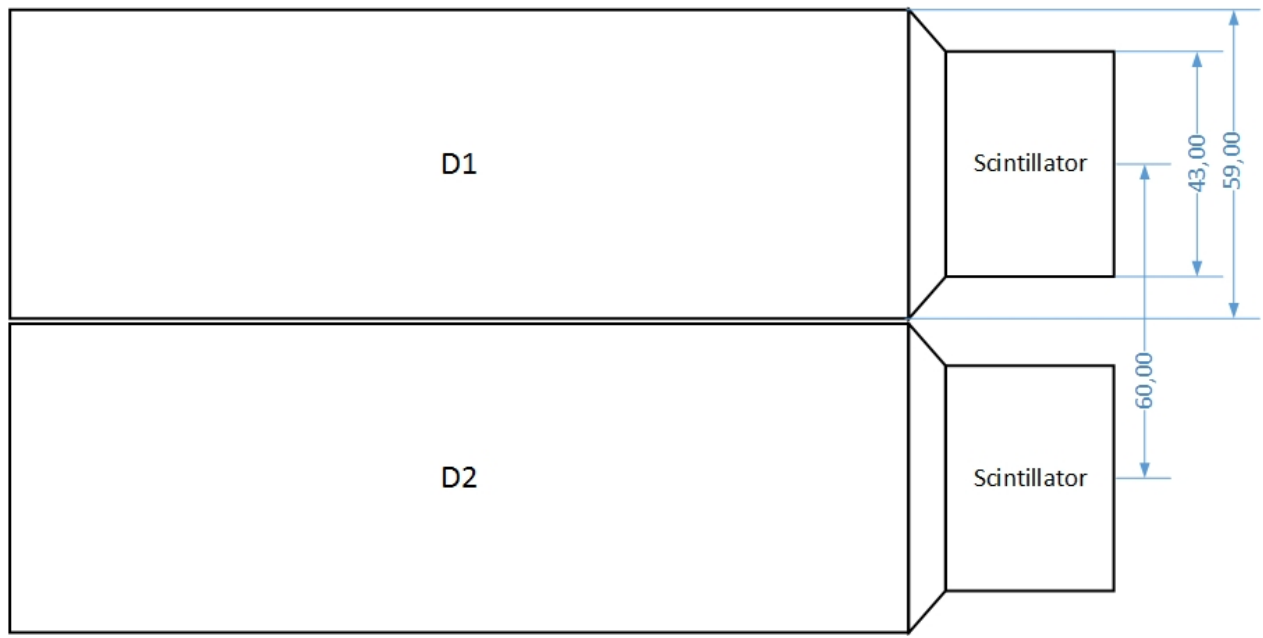

**Figure 5**. How the two cylindrical LaBr3 scintillation detectors are placed next to each other. The detectors register an X-ray signal simultaneously during some discharges. This proves that the photon flux is high enough to cause a multiphoton registration at one detector.

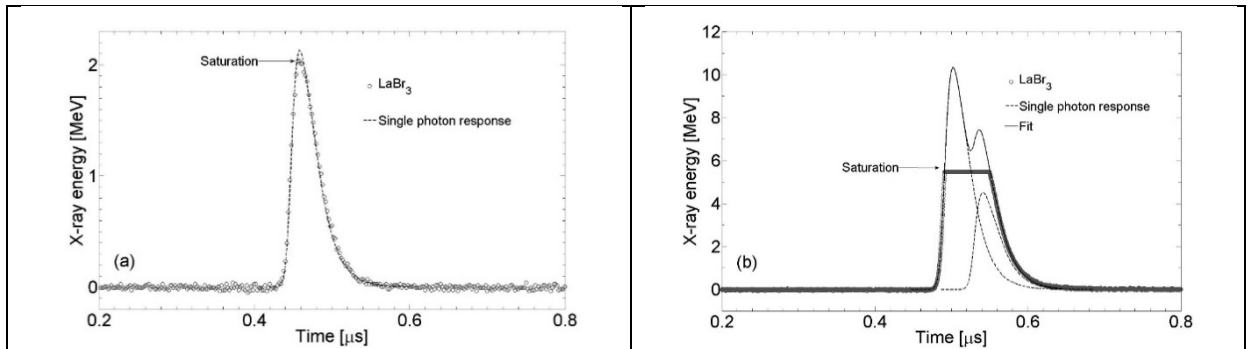

**Figure 6**. (a) An X-ray detector signal of 2 MeV. A single photon response fits the data well, while one would expect only multiple photons to generate a 2 MeV signal in a 1 MV discharge. At the peak the photomultiplier may be slightly saturated, or alternatively, the signal is a pile-up of two photons with 6 ns delay. (b) One of many possible fits to a measured signal. At least two X-ray bursts overlap, which leads to detector saturation. The signal of each burst, in turn, consists of multiple photons.

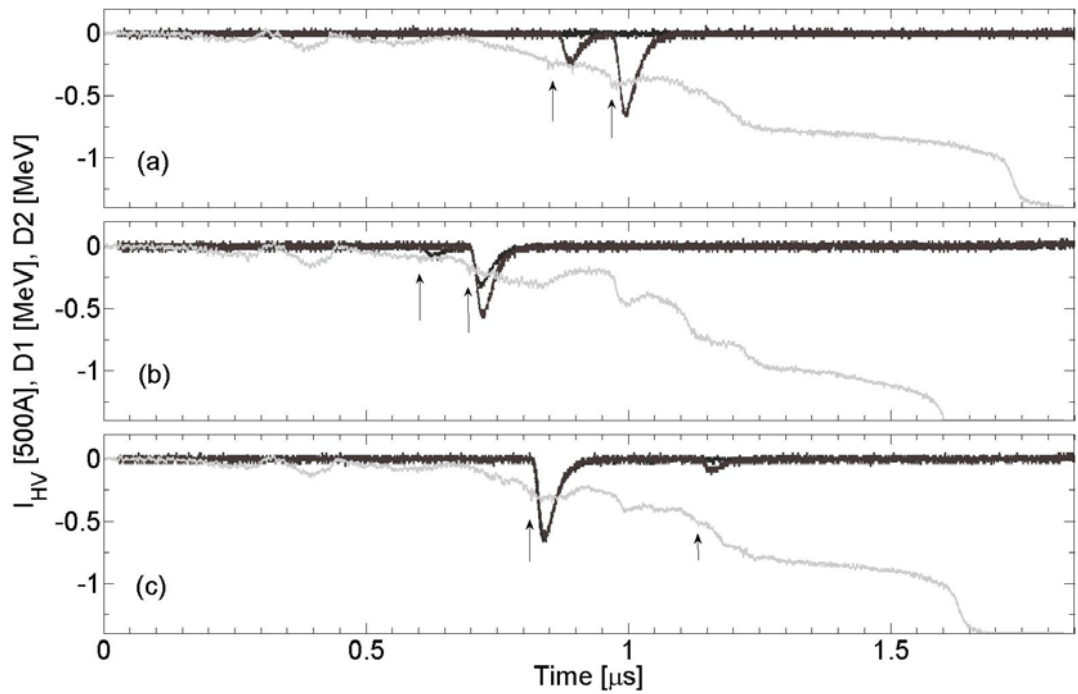

**Figure 7** (a)-(c): Three different discharges in a gap of 107 cm. Plotted are cathode current (gray) and X-ray detections by two detectors at position H as a function of time. In all three cases two separate X-ray bursts are detected. All X-ray bursts are accompanied by high frequency oscillations of the current (marked by arrows).

**Table 1.** The rate of discharges with X-rays detection for the different detector positions A to H (see Fig. 1). The effective detection area of one detector is 11.3 cm$^2$. The gap distance is 107 cm. The coordinate system is indicated in Figure 1.

| Detector position | Coordinates x; y (m) | Number of discharges with X-ray detection / number of discharges | Rate of discharges with X-ray detection (%) |
|---|---|---|---|
| A[1] | 0.15; -0.13 | 104/314 | 33 |
| B[1] | 0.35; -0.13 | 32/120 | 27 |
| C[1] | 2.10; 0.15 | 29/160 | 18 |
| D[1] | 1.50; 0.15 | 54/140 | 39 |
| E[2] | 1.15; -0.3 | 3/10 | 30 |
| F[2] | 1.50; 0.6 | 25/60 | 42 |
| F' up[2] | 1.50; 0.6 | 8/50 | 16 |
| F' down[2] | 1.50; 0.6 | 4/50 | 8 |
| G[2] | 1.50; 2.0 | 14/50 | 28 |
| H[3] | 2.10; 0.6 | 120/856 | 14 |
| [1] Series I, [2] Series II, [3] Series III | | | |

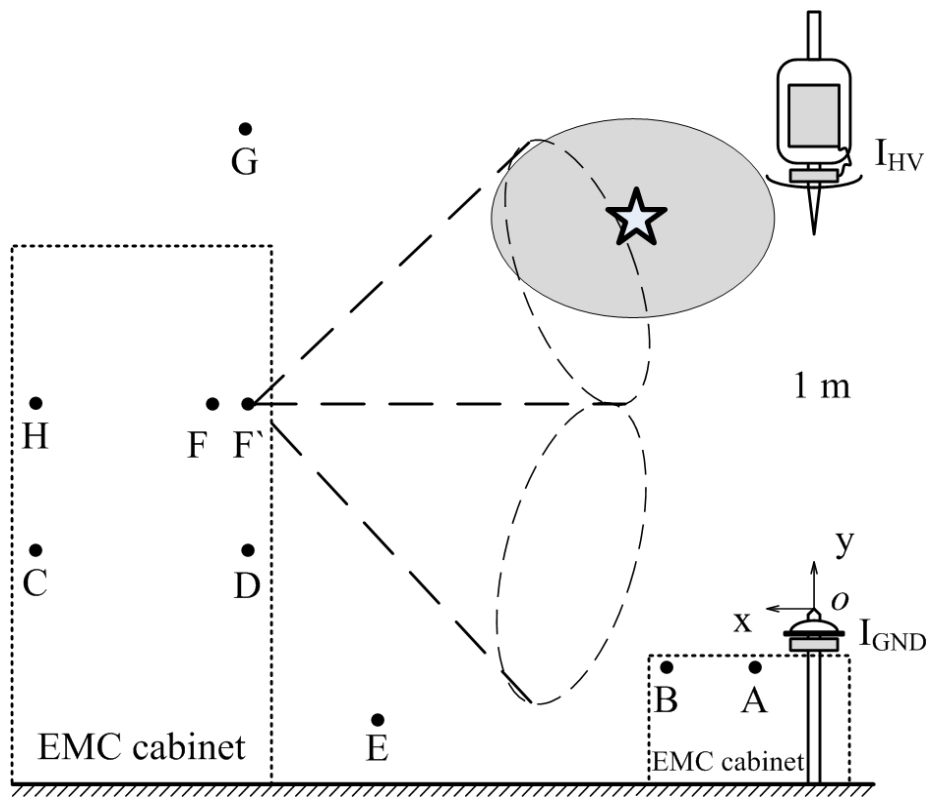

**Figure 8**. The source location with a confidence bound of 95% calculated with an inverse square law fit of the data shown in Table 1. Neither attenuation by air nor by detector/cabinet aluminum casings is taken into account. The approximate source location is near the HV electrode and off axis.

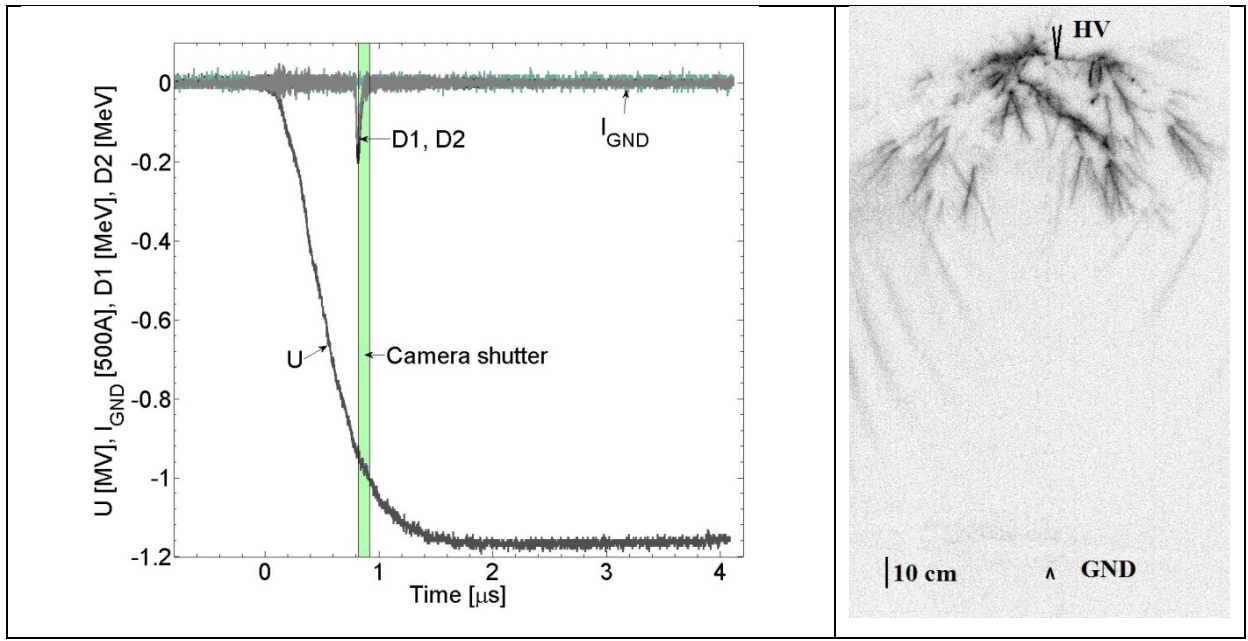

**Figure 9**. Here the distance between HV and GND electrodes is as large as 175 cm. X-rays are detected during the fourth streamer burst of the pre-breakdown phase. No current through the GND electrode is registered, and no light from the GND electrode detected. The discharge does not develop into a spark. The X-rays were detected by two LaBr3 detectors just before the picture was taken with an exposure time of 100 ns.

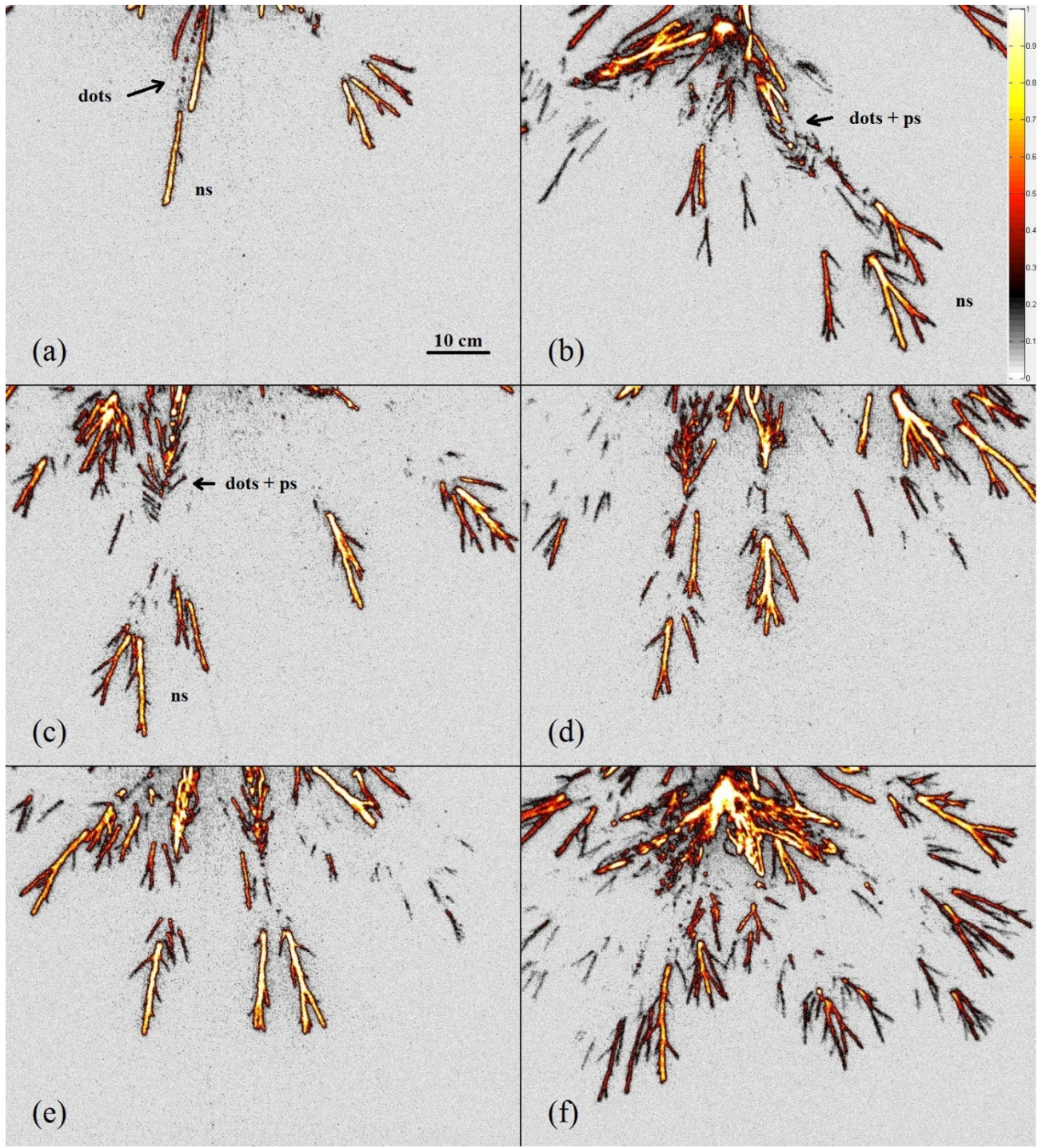

**Figure 10**. Images of the vicinity of the cathode at the time of X-ray detection with an exposure time of 50 ns. For all images the camera shutter is opened at the same time as in Figure 9 (b). (a) Negative streamers (ns) leave beads behind in pre-ionized medium. The dots act as starting points for new positive (ps) cathode-directed streamers (b) – (f). A possible collision between negative and positive streamers is visible in (c) and (d).

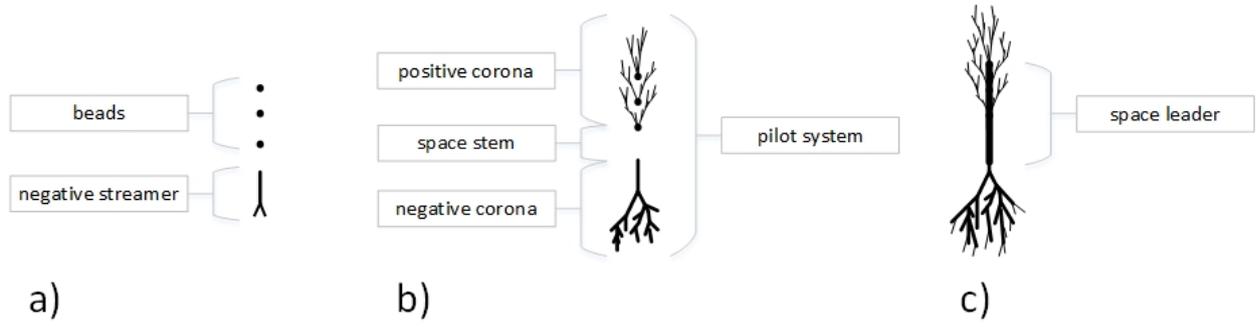

**Figure 11**. Schematic of the development of a pilot system. It is a bipolar structure that can develop into a space leader in sufficiently long gaps and apparently into lightning leaders.

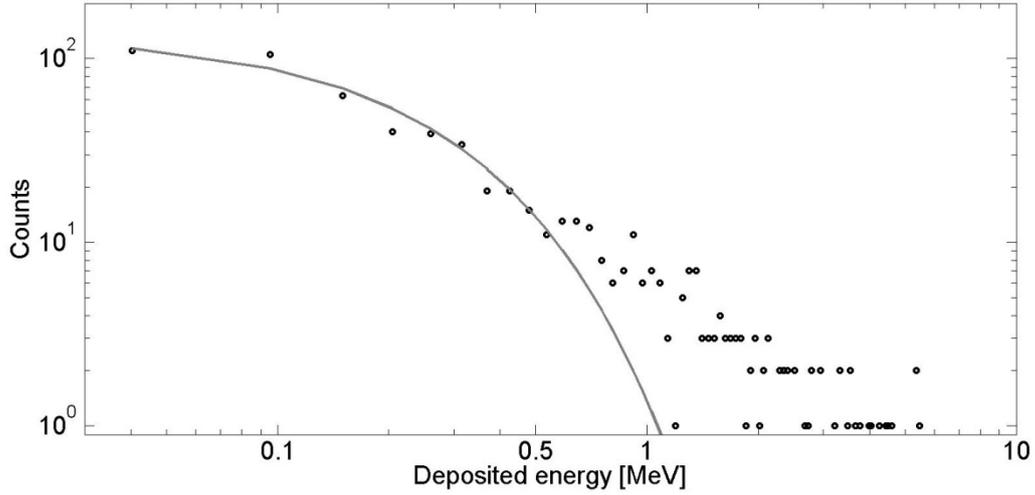

**Figure 12**. The pseudo-spectrum of X-rays (dots) collected by two LaBr$_3$ detectors at position H (Figure 1), indicating the energy deposition in a detector which also can be due to the pile-up of multiple photons in an X-ray burst. The energy bins are 55 keV wide, and the statistics is over 636 X-ray bursts. The solid line is a fit with $dn/d\varepsilon \sim \exp(-\varepsilon/\varepsilon_c)$, where $\varepsilon_c$ equals 0.2 MeV. While the low energetic part of spectrum fits well up to 0.5 MeV, the high-energetic part lies above the fit. This happens due to multiphoton counts within one X-ray burst, or due to the overlap of two or more X-ray bursts.

**Table 2**. The registration rate in % calculated from 50 discharges.

|         | Lead thickness, mm | | | | | |
|---------|----|-----|---|-----|---|-----|
|         | 0  | 1.5 | 3 | 4.5 | 6 | 7.5 |
| Point A | 31 | 6   | 2 | 0   | 0 | 0   |
| Point D | 32 | 8   | 5 | 0   | 0 | 2   |

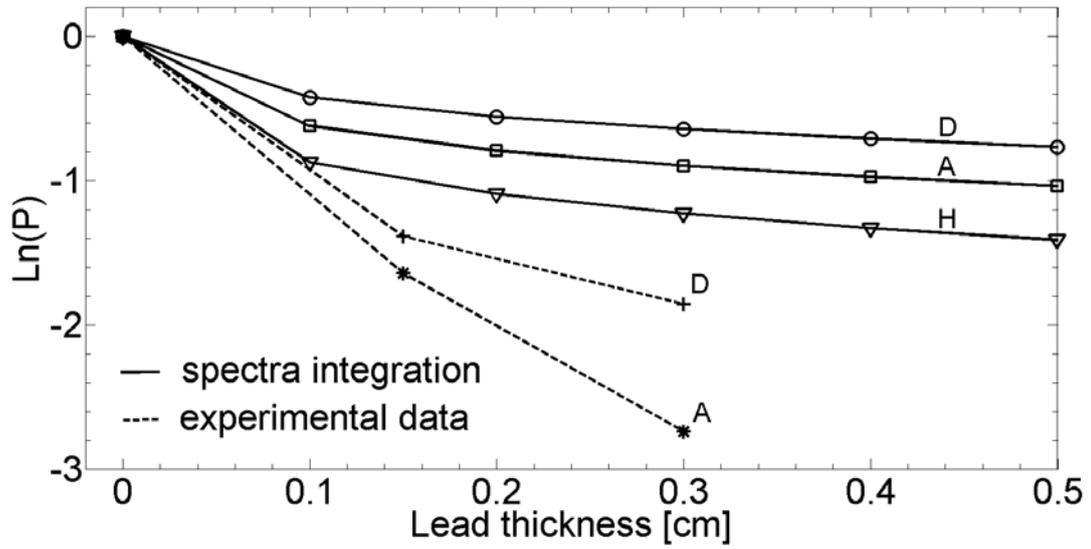

**Figure 13**. Experimental (dashed) and calculated (solid) attenuation curves at positions A, D and H. The calculated values are above the measured ones because of multiphoton registration and of overlapping X-ray bursts.